\documentclass[twocolumn,showpacs,preprintnumbers,amsmath,amssymb]{revtex4}

\usepackage{graphicx}
\usepackage{dcolumn}
\usepackage{bm}

\begin{document}


\title{Structural phase transitions of vortex matter in an optical lattice}
\author{H. Pu$^1$, L. O. Baksmaty$^2$, S. Yi$^1$, and N. P. Bigelow$^2$}

\affiliation{$^1$Department of Physics and Astronomy, and Rice
Quantum Institute, Rice University, Houston, TX 77251-1892\\
$^2$Department of Physics and Astronomy, and Laboratory for Laser
Energetics, University of Rochester, Rochester, NY 14627}

\begin{abstract}
We consider the vortex structure of a rapidly rotating trapped
atomic Bose-Einstein condensate in the presence of a co-rotating
periodic optical lattice potential. We observe a rich variety of
structural phases which reflect the interplay of the vortex-vortex
and vortex-lattice interactions. The lattice structure is very
sensitive to the ratio of vortices to pinning sites and we observe
structural phase transitions and domain formation as this ratio is
varied.
\end{abstract}

\date{\today}
\pacs{03.75.Mn, 03.75.Nt}

\maketitle

Structural phase transitions (SPTs) occur in a wide range of
physical and chemical systems~\cite{spt}. Studies of SPTs are both
of technological importance and of fundamental interest.
Technologically, an understanding of SPTs is essential to the
study of many materials; while fundamentally, SPTs are related to
a variety of questions in statistical mechanics, crystallography,
magnetism, surface science, to name a few. Among the most
intensively studied systems related to SPTs are graphite
intercalation compounds, niobates, adsorbed molecular monolayers
and vortex matter in type-$II$ superconductors. In this Letter we
introduce a new system for SPT study: that of vortex matter
created in a trapped Bose-Einstein condensate (BEC) formed in the
prescence of a co-rotating optical lattice (OL). This system has
the advantage of being experimentally realizable, tunable over a
wide range of interaction parameters, and describable by an
analytic {\em ab initio} theory.

We take inspiration from two recent developments in atomic BECs.
The first is the observation of vortex lattice excitations in
rapidly rotating BECs~\cite{jila}.   The second is the
set of proposals and demonstrations concerning periodic optical
lattice potentials created using holographic phase plates \cite{phase} or
amplitude masks \cite{mask}.  This latter technique means that
a rotating OL (at desired
angular frequency) can be realized by simply rotating the OL phase plates
or masks. The amenability of BECs to
imaging, experimental control and theoretical description makes the atomic condensate an attractive test
beds for many phenomena frequently encountered, but often
difficult to study in other systems. One such example is given by
the recent spectacular realization of the superfluid-Mott
insulator quantum phase transition in a condensate confined by an
optical lattice~\cite{mott} following a proposal of Jaksch {\em et
al.} \cite{zoller}. This transition was theoretically predicted
over a decade ago~\cite{mottcm} in the context of liquid $^4$He
absorbed in porous media.

In our system, the phase transition is observed as a shift in the
periodicity and symmetry of the vortex lattice as optical
potential parameters are tuned. Fundamentally, this situation is
analogous to that of a type-$II$ superconductor subject to
external magnetic field and artificial periodic
pinning~\cite{super,reichhardt,double}. In the analogy, the
angular rotation $\Omega \hat{z}$ of the BEC plays the role of the
magnetic field and the peaks of the OL play the role of the
pinning centers. In contrast to the pinning lattice in typical
superconducting samples, the periodicity and depth of the optical
lattice may be dynamically tuned in an atomic system. Furthermore,
the clean microscopic physics of atomic condensates makes a first
principles calculation possible based on a mean-field treatment.
This is in contrast to the superconductor case, where theoretical
calculations frequently rely on elegant, but phenomenological
models or molecular dynamics simulations \cite{reichhardt}.

We work in a pancake shaped geometry.  For the BEC, this is
justified because at high rotation rates, the centrifugal forces
reduce the radial trapping frequency $\omega_{\bot}$ and the
condensate may be accurately assumed to be frozen into the
harmonic oscillator ground state in axial direction ($\hat{z}$).
We therefore integrate the axial degree of freedom, obtaining an
effective two-dimensional system with a renormalized coupling
constant. In a frame rotating with angular velocity $\Omega
\hat{z}$ the transverse wave function $\phi (x,y)$ obeys the
two-dimensional time-dependent Gross-Pitaevskii equation:
\begin{equation}
\label{gpe} i \frac{\partial \phi}{\partial t} =
\left[-\frac{1}{2} \nabla^2 + \frac{r^2}{2} +V_{\rm lat}+U
|\phi|^2 -\mu- \Omega L_z \right] \phi.
\end{equation}
Unless otherwise specified we choose our units for time, length and energy as
$1/\omega_{\bot}$, $\sqrt{\hbar/(m \omega_{\bot})}$ and $\hbar \omega_{\bot}$,
respectively. Here $r^2=x^2+y^2$, $U$ is the (renormalized) effective
2D nonlinear interaction
coefficient, $\mu$ the chemical potential, $L_z$ the $z$-component
of the angular momentum operator, and $V_{\rm lat}= V_0 [\sin^2(k
x) + \sin^2 (ky)]$ is the optical lattice potential. In this work,
we only consider the case where the optical lattice is co-rotating
with the condensate, thus $V_{\rm lat}$ in Eq.~(\ref{gpe}) is
time-independent.

\begin{figure*}
\includegraphics*[width=16cm,height=11.cm]{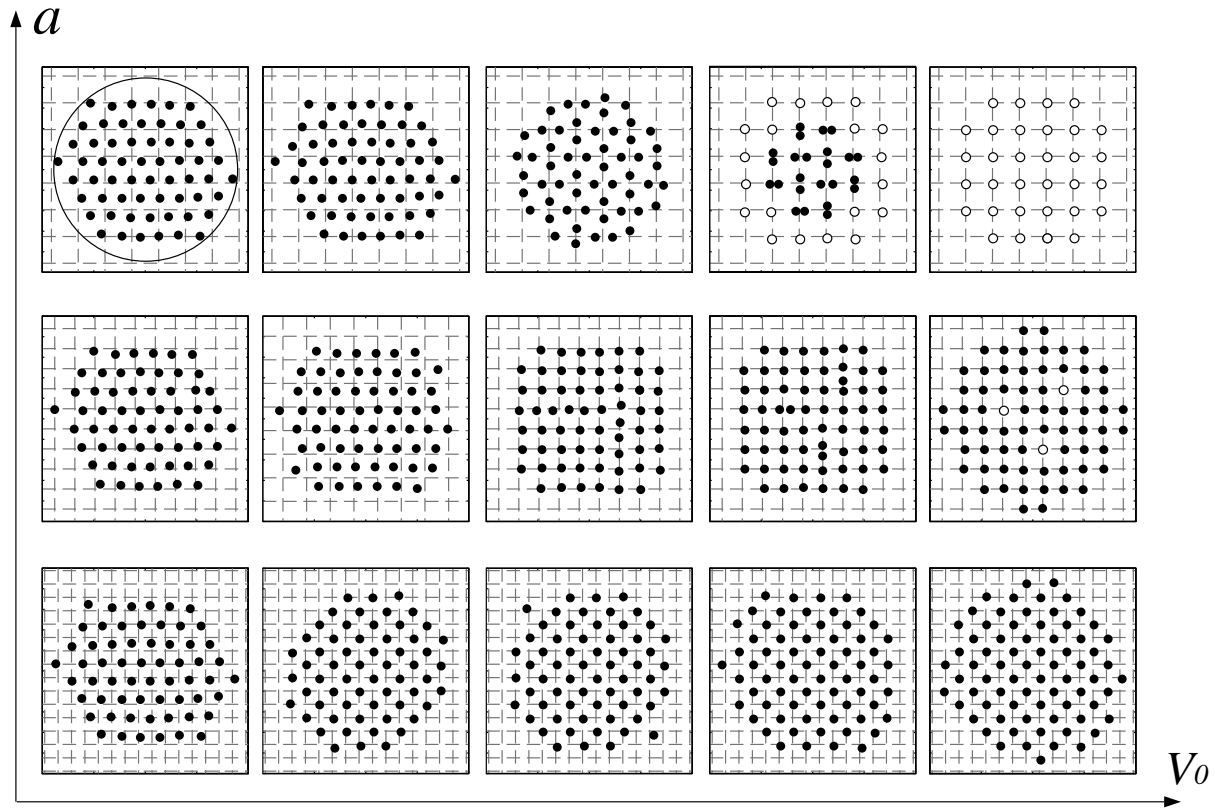}
\caption{Structure of the ground state. Each intersection of the
dashed lines represents an antinode of the optical potential.
Filled ($\bullet$) and empty ($\circ$) circles represent the
positions of singly and doubly quantized vortices, respectively.
Only the positions of vortices within the Thomas-Fermi radius are
shown. The Thomas-Fermi boundary $R_{TF}=\sqrt{2\mu}$ is
represented by the circle in the upper-left plot. The upper,
middle and lower rows correspond to $a=4 \epsilon/3$, $\epsilon$
and $2 \epsilon/3$, or $\eta \approx$2, 1.15 and 1/2,
respectively. From left to right, $V_0=0.05$, 0.2, 0.5, 2.0 and
5.0, respectively. } \label{fig1}
\end{figure*}

For each chosen value of the OL periodicity (denoted by
$a=\pi/k$), we vary the peak-to-peak potential amplitude of the
optical lattice $V_0$ from 0 to $V_{\rm max}$, a value at which
the lattice is fully pinned. We obtain the ground state of the
system for a given set of parameters ($V_0,a,\Omega$) by
propagating Eq.~\ref{gpe} in imaginary time (i.e. by steepest
descent). To ensure that the obtained structure corresponds to the
ground state, we usually start with several different initial
trial wave functions possessing different symmetries for each set
of parameters. In our calculation, we use the $^{87}$Rb parameters
of the JILA experiment \cite{jila}: $\omega_{\bot}=2\pi \times
8.3$Hz, $\omega_{z}=2\pi \times 5.3$Hz, $\Omega = 0.95
\omega_{\bot}$, and the total particle number is taken to be $2
\times 10^5$. Evidently, both the atomic and optical parameters
will affect the ground state. For simplicity, in this work we
focus on the effect resulting from variation of the OL. Our variables are
therefore the depth ($V_0$) and periodicity ($a$) of the optical
potential. Specifically, $a$ defines the density of pinning sites
which occur at the peaks of the OL potential whereas $\Omega$
determines the density of vortices. As found in previous
studies~\cite{reichhardt}, we observe that the structure of the
fully pinned vortex lattice (FPVL) is very sensitive to the ratio
$\eta$ defined as $\eta=N_{v}/N_{p}$, where $N_{v}$ and $N_{p}$
are the density of vortices and pinning sites, respectively. For
the OL potential we use here, $N_p=1/a^2$. In the absence of the
OL, $N_v=2/(\sqrt{3}\epsilon^2)$, where
$\epsilon=\sqrt{2\pi/(\sqrt{3} \Omega)}$ is the inter-vortex
spacing in a completely unpinned ($V_0=0$) vortex
lattice~\cite{fetter}. This value of $N_v$ is not significantly
changed by the presence of the OL. Hence we take $\eta \approx
2a^2/(\sqrt{3}\epsilon^2)$. In brief, we find
that when $\eta \approx 1/2$, the FPVL has a checkerboard
structure and when $\eta \approx 1$, we obtain a square lattice
while when $\eta \approx 2$ we obtain a square lattice of doubly
charged vortices.

The trends in our results are graphically illustrated in
Fig.~\ref{fig1}, our main result. For three values of $a$ the
ground state vortex structure is plotted for increasing values of
$V_0$. The upper, middle and lower rows correspond to different
pinning site densities represented by $a=4 \epsilon/3$, $\epsilon$
and $2 \epsilon/3$, respectively. In this situation, the angular
momentum of the condensate is carried in singly quantized vortices
which organize into a triangular Abrikosov lattice due to the
logarithmically repulsive inter-vortex interaction. Such vortex
lattices have been observed~\cite{vl}, perturbed~\cite{jila} and
accurately described~\cite{anglin} by several groups in recent
years.

As is not surprising, for sufficiently small $V_0$ the vortex
lattice maintains a triangular geometry with only slight
distortions (see first column of Fig.~\ref{fig1}). In the opposite
extreme, at sufficiently large values of $V_0$, all the vortices
are pinned to the antinodes of the OL potential, mirroring the
geometry set by the OL potential.  What is remarkable are the
states that exist between these extremes.  At $a=4 \epsilon/3$
(upper row of Fig.~\ref{fig1}), there are about twice as many
vortices as pinning sites ($\eta \approx 2$), and the FPVL is
found to be a square lattice of doubly quantized vortices which is
commensurate with the OL, a situation analogous to the second
matching field case for the superconductor system in which the
observation of a square double-quanta vortex lattice was reported
in Ref.~\cite{double}. At a higher pinning site density defined by
$a= \epsilon$ (middle row of Fig.~\ref{fig1}, $\eta \approx
1.15$), most of the antinodes of the OL are occupied by singly
quantized vortices with the exception of three doubly quantized
vortices. And finally, at $a=2 \epsilon/3$ (lower row of
Fig.~\ref{fig1}, $\eta \approx 1/2$), every next-nearest-neighbor
site of the OL is occupied by a singly quantized vortex and the
FPVL forms what can be described as a ``checkerboard'' square
lattice rotated $45^\circ$ with respect to the OL, once again
analogous to the half matching field case of the superconductor
system \cite{reichhardt}.

For intermediate values of $V_0$, the vortices form structures in
between the triangular lattice and the FPVL, the details of which
also depend on the period of the OL. For example, at $a=4
\epsilon/3$, for which the FPVL is a square lattice of doubly
charged vortices, we observe bound pairs centered around the OL
pinning sites for the potential depth in the range $0.5 < V_0 <
4.0$. We point out that the orientation of each pair is orthogonal
to all adjacent ones. As $V_0$ increases, these pairs are more and
more tightly bound and eventually all pairs collapse onto the
corresponding pinning sites thus forming doubly quantized
vortices. Further increasing the OL period, vortices with higher
and higher winding number will start to appear in the FPVL.

\begin{figure}
\includegraphics*[width=8cm,height=7.5cm]{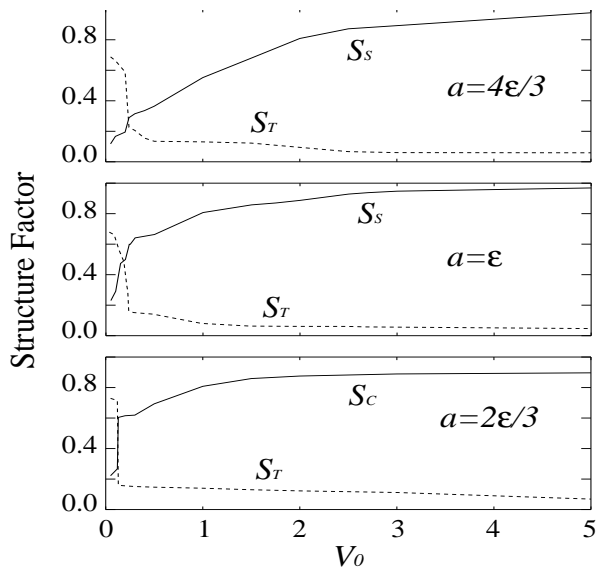}
\caption{Structure factors as functions of $V_0$ for
$a=4\epsilon/3$, $\epsilon$ and $2\epsilon/3$. $S_{T,S,C}=|S({\bf
k}_{T,S,C})|$, where ${\bf k}_{T,S,C}=(2\pi/\epsilon)\hat{x} -
[2\pi/(\sqrt{3}\epsilon)] \hat{y}$, $2k\hat{y}$ and $k \hat{x}-k
\hat{y}$, representing respectively one of the two fundamental
reciprocal vectors for the triangular, square and checkerboard
lattices. } \label{fig2}
\end{figure}

In order to characterize the SPT more quantitatively, we calculate the
structure factor~\cite{reichhardt} of the vortex lattice defined as:
\begin{equation}
S({\bf k})  = \frac{1}{N_c} \sum_i n_i\, e^{i {\bf k} \cdot {\bf
r}_i}  , \label{sk}
\end{equation}
where $i$ labels individual vortices, $n_i$ and ${\bf r}_i$ are
the winding number and position of the $i$-th vortex, while $N_c
=\sum_i n_i$ is the total winding number. For a familiar crystal
lattice, $|S({\bf k})|$ displays peaks at the corresponding
reciprocal lattice vectors. Here, we focus on following three
cases: (1) the triangular Abrikosov lattice in the absence of the
OL ($S_{T}$), (2) the square ($S_{S}$) and (3) the checkerboard
($S_{C}$) lattices defined by the OL. Each lattice structure has
two fundamental reciprocal vectors. In this instance it is
sufficient to calculate the structure factor along one dimension,
i.e., for one of the reciprocal vectors.  Our results are
displayed in Fig.~\ref{fig2}. We observe that as $V_0$ is
increased from zero, the triangular lattice is destroyed over a
very small range of $V_0$ as $S_T$ exhibits a sudden jump,
indicative of a first order transition which is physically
expressed by the motion of vortices towards the pinning sites.
When there are more vortices than pinning sites ($\eta > 1$), the
surplus vortices get pinned at a comparatively ``slow'' pace as a
consequence of repulsion experienced from vortices which are
already pinned. Conversely, when the number of pinning sites
exceeds that of vortices ($\eta < 1$), the FPVL is quickly
established.

\begin{figure*}
\includegraphics*[width=16cm,height=3.9cm]{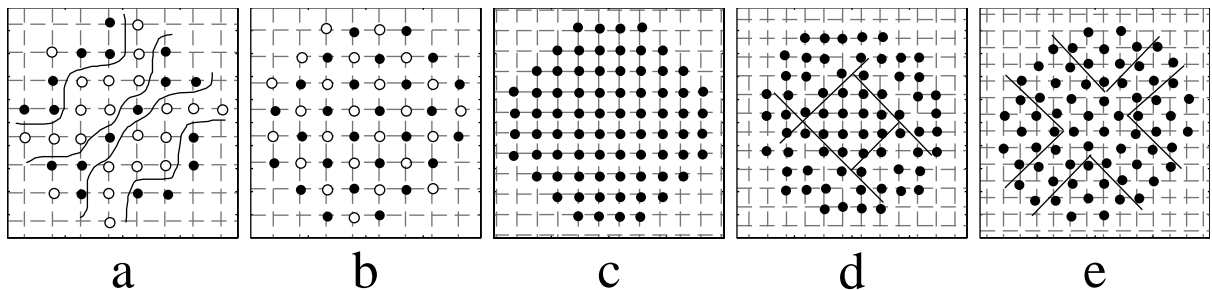}
\caption{Structure of the FPVL. From left to right, $\eta=$1.8,
1.6, 1.0, 0.8 and 0.6, corresponding to $a/\epsilon=$1.25, 1.176,
0.95, 0.833 and 0.741, respectively. Solid lines represent domain
walls.} \label{fig3}
\end{figure*}

We now turn to explore the structure of the FPVL for various
values of $\eta$. In the case $\eta$ is not close to an integer or
inverse integer value, we observe the coexistence of sub-lattices
of different geometry bounded by domain walls which always lie
along the diagonal of the OL. For $\eta=1.8$ we have the
coexistence of two sub-lattices formed by doubly and singly
quantized vortices, respectively (see Fig.~\ref{fig3}a). For this
FPVL the domains have striped parallel walls (represented by the
solid lines in the figure), indicating inter-domain
repulsion~\cite{wall1,wall2}. This can be understood as a
consequence the repulsive vortex-vortex interaction. For the
slightly smaller value of $\eta=1.6$, these two sub-lattices form
two interlocking checkerboard structure (see Fig.~\ref{fig3}b). As
$a$ decreases further to 0.95$\epsilon$ (Fig.~\ref{fig3}c), all
doubly quantized vortices have disappeared and the vortex lattice
becomes completely commensurate with the OL, with each pinning
site hosting a singly quantized vortex. This represents the fully
matching case of $\eta=1$. Upon further decrease of $a$ or $\eta$,
Fig.~\ref{fig3}d and e show that more pinning sites become
unoccupied and the checkerboard domain begins to cover the whole
lattice. In this ground state, the walls between the square and
checkerboard domains are crossing each other, signifying an
attractive domain wall interaction which may be intuitively
understood as resulting from the tendency of the vortices to
occupy the vacant pinning sites.

In the study of domain wall formation, we see again distinct advantages of
 BEC vortex lattices over alternative systems. Although
similar domain formation was observed earlier in superconducting
systems~\cite{double,wallexp}, their origin could not be clearly
established because of additional defects likely to be present in
the experimental samples. By contrast, in the atomic BEC system,
the OL provides us with a defect-free periodic pinning potential
and thus frees us to confidently investigate the more fundamental
factors controlling the dynamics of domain formation and hence the
structure of the domain walls. Recent success at directly
imaging~\cite{jila} and calculating~\cite{anglin} vortex lattice
excitations in BECs presents very exciting possibilities for
studying, in unprecedented detail, the dynamics of the FPVLs
obtained in this work and opens the door to more ambitious studies
of SPTs in unconventional geometries.

In summary, we have theoretically investigated the vortex state of
a rapidly rotating condensate in a periodic optical potential. We
have found that the vortex lattice exhibits a rich variety of
structures depending on the parameters of the optical potential.
In the future it will be interesting to investigate the detailed
dynamics (e.g. time evolution) of the phase transitions between
various structures and to relate the transitions to the properties
of the condensate such as its collective excitation modes. Such
studies will certainly shed new light on many other systems
displaying structural phase transitions. In the current work, we
have assumed that the optical potential is co-rotating with the
condensate. We also plan to generalize this work to the case when
the OL and the condensate are not co-rotating, under which
condition, dynamical phases and phase transitions may exist
\cite{dyna}.

This work is supported by the NSF, the ONR and the University of
Rochester (LOB and NPB) and by Rice University (HP and SY). LOB is
supported by the Laboratory for Laser Energetics Horton Program.
NPB is also with the Institute of Optics.  We thank Ennio Arimondo
for fruitful discussions.

{\em Note added}---During the preparation of the paper, we noticed
the work of Reijnders and Duine \cite{duine} who studied some
aspects of this system using a variational approach.

\end{document}